\documentstyle[12pt]{article}

\newcommand{\F}{\noindent}

\newcommand{\MP}{\medskip}
\newcommand{\BP}{\bigskip}

\newcommand{\HH}{{\cal H}}

\newcommand{\beq}{\begin{eqnarray}}
\newcommand{\ene}{\end{eqnarray}}

\Large

\begin{document}
\begin{flushright}
KIMS-1994-08-20\\
gr-qc/0110065
\end{flushright}

\BP

\begin{center}
\Large

{\bf Local Time and the Unification of Physics}

{\bf Part I. Local Time}

\normalsize

\vskip12pt

Hitoshi Kitada

Department of Mathematical Sciences, University of
Tokyo

Komaba, Meguro, Tokyo 153, Japan

E-mail: kitada@po.iijnet.or.jp
\vskip6pt

and

\vskip6pt

Lancelot R. Fletcher

30-3406 Newport Parkway

Jersey City, New Jersey 07310, USA

E-mail: lance@freelance.com
\vskip10pt

April 8, 1996
\vskip2pt

(published in {\it Apeiron}, vol.3, no.2, April 1996, pp.38--45)

\end{center}

\vskip24pt

\normalsize

\leftskip24pt
\rightskip24pt

\small

\noindent
{\it Abstract}:
The notions of time in the theories of Newton and Einstein are
 reviewed so that the difficulty which impedes the unification of
 quantum mechanics (QM) and general relativity (GR) is clarified. It
 is seen that GR by itself contains an intrinsic difficulty relating
 to the definition of local clocks, as well as that GR still
 requires a kind of absolute that can serve as an objective
 reference standard. We present a new understanding of time, which
 gives a consistent definition of a local time associated with each
 local system in a quantum mechanical way, so that it serves the
 requirements of both GR as well as QM. As a consequence, QM and GR
 are reconciled while preserving the current mathematical
 formulations of both theories.

\leftskip0pt
\rightskip0pt

\vskip 24pt

\large
\noindent
{\bf I. Introduction}

\vskip12pt

\normalsize

\noindent
Previous papers of Kitada 1994a, 1994b proposed an approach to
the problem of overcoming the apparent inconsistency of
non-relativistic quantum mechanics and general relativity. The
purpose of this paper is to explain the structure and background
of that approach, with emphasis on a certain philosophical problem
related with the notion of time.

The inconsistency of quantum mechanics
and general relativity, when
looked at mathematically, seems at first sight obvious and
inescapable from the fact that the geometry of quantum mechanics
is Euclidean, while general relativity employs a curved,
Riemannian geometry.

Kitada 1994a proposes to overcome this
apparent mathematical incommensurability of these two geometries
by ``orthogonalizing" them; {\it i.e.} by expressing them as a direct
product $X\times R^6$, where $X$ represents the curved
Riemannian manifold associated with general relativity, and
$R^6$ (or in the usual space-time context, $R^4$) denotes the
Euclidean space of phase space coordinates $(x,v)$ of
non-relativistic quantum mechanics. As two components of the
orthogonalized total space $X\times R^6$, the Riemannian space
$X$ and the Euclidean space $R^6$ are compatible without
contradiction.

General relativity and quantum mechanics are the two most
important and comprehensive theories of contemporary physics. By
``comprehensive" we mean that both theories claim to apply to
everything. In practice it might seem that these two theories
describe two different physical domains, since the most striking
applications of quantum mechanics occur when we consider things
that are extremely tiny in relation to ourselves -- things like
electrons and photons -- while the most striking applications of
general relativity occur in connection with extremely large and
dense concentrations of matter and enormous spatio-temporal
magnitudes. But, in principle, every physical thing must be
capable of being described adequately by both theories, at
least this is what the theories claim. And there are certain
cases -- of particular interest in recent cosmology and
astrophysics -- where the extremes of density that are the
particular province of general relativity coincide with the
extremes of minuteness that are the special province of quantum
mechanics. In those situations, the physicist is compelled to
face a problem which is present in the background of science all
the time but which can otherwise be evaded without practical
consequence: the fact, namely, that these two comprehensive
theoretical structures appear to be mutually incompatible, that
they seem to involve different -- and contradictory --
assumptions about the nature of space, time and causality.

Our intention in this paper is to outline an approach to the
understanding of general relativity and quantum mechanics in
which these theories will appear as distinct but systematically
coordinated perspectives on the same reality. The
orthogonalization of the spatial foundations of the two theories
allows us to speak of the two theories as distinct. To express
the possibility of their systematic coordination will require a
more extended analysis of the nature of time.

In brief, we can express our approach as follows:

\begin{enumerate}
	
\item We begin by distinguishing the notion of a local system
consisting of a finite number of particles.
Here we mean by ``local" that
the positions of all particles in a local system are understood
as defined with respect to the same reference frame.

\item In so far as the particles comprised in this local system are
understood locally, we note that these particles are describable
only in terms of quantum mechanics. In other words, to the
extent that we consider the particles solely within the local
reference frame, these particles have only
quantum mechanical properties, and cannot be described as
classical particles in accordance with general relativity.

\item Next we consider the center of mass of a local system.
Although the local system is considered as composed of particles
which -- as local -- have only quantum mechanical properties, in
our orthogonal approach we posit that each point $(t,x)$ in the
Riemannian manifold $X$ is correlated to the center of mass of
some local system. Therefore, in our approach, the classical
particles whose behavior is described by the general theory of
relativity are {\it not} understood as identical with the ``quantum
mechanical" particles inhabiting the local system -- rather the
classical particles are understood as precisely correlated only
with the centers of mass of the local systems.

\item It is important to recognize that the distinction we are
making between local systems and classical particles which are
the centers of mass of local systems is not a simple distinction
of inclusion/exclusion. For example, we may consider a local
system containing some set of particles, and within that set of
particles we may identify a number of subordinate ``sublocal"
systems. It would seem that the centers of mass of these
sublocal systems must be ``inside" the local system as originally
defined, but the sublocal system is at the same time a local
system, and we have said that the centers of mass of local
systems	are correlated with classical particles whose behavior
is to be described in terms of relativity theory.

\end{enumerate}

\F
The paradox is avoided by noting that the distinction we are
making is a distinction of reference frame, not a distinction of
inclusion or exclusion. When we speak of classical particles (or
centers of mass) we are speaking of the particle in terms of the
observer's time, which is understood as distinct from that of
the particle observed. To the extent that the time of the system
$L$ itself is adopted as the reference time, then we are
speaking of the behavior of a local system whose development
must be described in terms of quantum mechanics{\footnote{The
mathematical details of the relationship between the local
 time and the observer's time will be set forth in Part II,
 after we have developed our notion of local time in
 section IV of this part.}}.

It is our contention that time necessarily has two quite
different aspects, in relativity theory, on the one hand, and in
quantum theory on the other, and the intention of this paper is
to show that these two aspects of time are in fact complementary
and that the notion of local time, which we have associated with
the quantum mechanical local system, is not only the main
ingredient of a unification of quantum and relativity theories,
but that this actually is necessary to constituting the time of
relativity theory.

The ``orthogonalization" of the geometries of quantum mechanics
and general relativity does not by itself specify the nature of the
relationship between them. It simply gives us a way of representing
them as independent but complementary. The nature of that
relationship, and the value of this form of representation, will
come to light in Part II, after we outline our notion of local time
below.

Before stating our notion of local time, it will be useful to show
how this notion relates to the apparent inconsistency of quantum
mechanics and general relativity.

\vskip22pt

\large
\noindent
{\bf II. Time in Quantum Mechanics}

\vskip12pt

\normalsize

\noindent
That the difficulty of reconciling quantum mechanics and general
relativity is connected to the question of time is now generally
recognized (see especially Isham 1993, Unruh 1993 and Hartle 1993).
What is central is the divergent relationship of these two branches
of modern physics to their common Newtonian heritage.

At the beginning of modern physics, Isaac Newton specified his notion of
time in the {\it Principia} as follows (Newton 1962, p.6):

\begin{quotation}

\F
Absolute, true, and mathematical time, of itself, and from its own nature,
 flows equably without relation to anything external, and by another name
 is called duration: relative, apparent, and common time, is some sensible
 and external (whether accurate or unequable) measure of duration by
 the means of motion, which is commonly used instead of true time; such
 as an hour, a day, a month, a year.

\end{quotation}

\F
Also in pp.7-8, he states:

\begin{quotation}

Absolute time, in astronomy, is distinguished
 from relative, by the
equation or correction of the apparent time.
For the natural days are truly
unequal, though they are commonly considered as equal,
 and used for a
measure of time; astronomers correct
this inequality that they may measure
the celestial motions by a more accurate time.
 It may be, that there is
 no such thing as an equable motion,
 whereby time may be accurately measured.
 All motions may be accelerated and retarded,
 but the flowing of absolute
 time is not liable to any change. The duration or
 perseverance of the
 existence of things remains the same,
 whether the motions are swift or slow,
or none at all: and therefore this duration ought
 to be distinguished from
 what are only sensible measures thereof;
 and from which we deduce it,
 by means of the astronomical equation.
The necessity of this equation, for determining
 the times of a phenomenon,
 is evinced as well from the experiments of the pendulum clock, as
 by eclipses of the satellites of Jupiter.

\end{quotation}

\noindent
The main point of this famous
passage is to assert the existence of an absolute, true time.
However, it is important to note that Newton asserts the
existence of his absolute time by means of a distinction. There
is absolute time, which flows without reference to anything
external, and then there is relative, apparent, or common time,
which is a measure of duration made by comparison of motions.
Not only that, but although there may be no absolutely regular
motion by means of which absolute time may be accurately
represented, absolute time is an ideal standard by means of
which relative or common time is ``corrected."

Einstein's theory of relativity, as is well-known, sharply
contrasts with Newton precisely on the question of time and
space: Einstein's theory makes no reference to either absolute
time or absolute space. Einstein retains the relative or common
time which can be measured and determined by means of actual
clocks associated with each local observer, but he completely
jettisons Newton's notion of an absolute time flowing equably
for all observers.

But precisely in this respect, quantum mechanics stands in sharp
contrast to relativity -- especially the general theory of
relativity. In quantum mechanics, unlike relativity, the time
parameter continues to be treated in an essentially Newtonian manner.

That time plays a
special, absolute role in quantum mechanics is evident in
Schr\"odinger equation (at least as customarily interpreted):
$$
\frac{\hbar}{i} \frac{\partial}{\partial t}\psi(x,t)+H\psi(x,t)=0,\quad
\psi(x,0)=\psi_0(x),
$$
where the Schr\"odinger operator or the Hamiltonian $H$ of the system
 is defined by
$$
H\psi(x,t)=-\frac{\hbar^2}{2m}
\sum_{j=1}^3\frac{\partial^2\psi}{\partial x_j^2}(x,t)
+V(x)\psi(x,t).
$$
Thus the solution of the Schr\"odinger equation is given by
$$
\psi(x,t)=\exp[-itH/\hbar]\psi_0.
$$
In this context, the time $t$ appears to be
given {\it a priori}, and then the motion
$\psi(x,t)$ of the system is derived from the Schr\"odinger
 equation by using the time evolution  $\exp[-itH/\hbar]$ of the
 system{\footnote{See Kitada
1994a and sections IV and V below for a different interpretation
of this solution, which will supply the key to our notion of a
`local clock'.}}.

 The fact that time
plays a special role in quantum mechanics may also be seen by
looking
 at the alternative formulation of quantum mechanics by Feynman 1948.
See also Kitada 1980 for the relation between the classical mechanics
and quantum mechanics researched along the line given by Feynman 1948.

Because in non-relativistic quantum mechanics the
 time-evolution of a system
is governed by
the Schr\"odinger equation, 
space and time in quantum mechanics are intrinsically
Newtonian
 in the sense that the form of Schr\"odinger equation
is not invariant with respect to the relativistic transformation
of coordinates.

\vskip24pt

\large
\noindent
{\bf III. What would an adequate notion of time require?}
\normalsize

\vskip12pt

\noindent
We have indicated that the primary source of the inconsistency
between QM and GR is to be found in their divergent and apparently
incompatible ways of treating time. Einstein pointedly rejects
Newton's idea of absolute time and treats time as something which is
locally defined by means of clocks which are at the same time
physical objects. In non-relativistic QM, on the other hand, the
role of time is essentially Newtonian, in the sense that time is an
external, background parameter.

Newton's statements about absolute time and space were controversial
from the time they were first published, and Einstein was by no
means the first to call them into question. Einstein seems to have
considered the Newtonian absolutes as purely metaphysical in nature,
having no direct bearing on actual physical description -- even for
classical mechanics. In his own presentations Einstein is
consistently operational. The description of spatial relations,
prior to the introduction of GR, is explained as involving the
specification of places on rigid reference bodies and spatial
co-ordinate systems are understood as convenient, abstract
mathematical substitutes for such rigid bodies of reference. Time
is understood, not as some absolute parameter, but as something
based on the readings of identically constructed clocks held in the
hands of different observers, who match up the ``ticks" of their
clocks with the observed positions of the various objects which they
are observing.

In short, it looks as if Einstein considered the Newtonian absolutes
completely superfluous and thought that they could be disposed of
with no consequence, and in fact much gain in scientific rigor.
However, once we inquire into the epistemological function of the
Newtonian absolutes we may discover that it is not in fact quite so
easy to get rid of them without any consequence.

Clearly, the epistemological function of the Newtonian absolutes is
to serve as a common reference standard. The idea is that there must
be something which is the same for all observers, in terms of which
all can be described. To that extent, it is clear that in the
physics of Einstein, Newton's absolutes have not actually been
removed -- they have only been disguised. So the notions of a rigid
body and a ``standard clock" are essentially surrogates for absolute
space and time. This can be seen to some extent by comparing Weyl's
definition of a clock with Newton's definition of absolute time:

Weyl 1952 (p.7) has this to say on the question of clocks and the 
measurement
of time:
\begin{quote}
	To be able to apply mathematical conceptions to questions of
	Time we must postulate that it is theoretically possible to fix
	in Time, to any order of accuracy, an absolutely rigorous {\bf now}
	(present) as a {\bf point of Time} -- i.e. to be able to indicate
	points of time, one of which will always be the earlier and the
	other the later. The following principle will hold for this
	``order-relation". If A is earlier than B and B is earlier than
	C, then A is earlier than C. Each two points of Time, A and B,
	of which A is the earlier, mark off a {\bf length of time}; this
	includes every point which is later than A and earlier than B.
	The fact that Time is a form of our stream of experience is
	expressed in the idea of {\bf equality}: the empirical content which
	fills the length of Time AB can in itself be put into any other
	time without being in any way different from what it is. The
	length of time which it would then occupy is equal to the
	distance AB. This, with the help of the principle of causality,
	gives us the following objective criterion in physics for equal
	lengths of time. If an absolutely isolated physical system (i.e.
	one not subject to external influences) reverts once again to
	exactly the same state as that in which it was at some earlier
	instant, then the same succession of states will be repeated in
	time and the whole series of events will constitute a cycle. In
	general such a system is called a {\bf clock}. Each period of the
	cycle then lasts {\bf equally} long.
\end{quote}
In particular note his reference to ``...an absolutely isolated
physical system (i.e. one not subject to external influences)," and
compare it to Newton's statement that, ``Absolute, true, and
mathematical time, of itself, and from its own nature, flows equably
without relation to anything external..." Clearly, Newton's absolute
time is unobservable -- and from the standpoint of empirical science
this is a serious fault which makes it unfit to serve as a standard
of measurement. But in fact Weyl's clock -- defined as an
absolutely isolated physical system -- suffers from exactly the same
problem: that which allows it to be accurate makes it at the same
time unobservable. Indeed, strictly speaking, Weyl's clock must be
exactly the same thing as Newton's, since the only absolutely
isolated physical system one can imagine is the universe itself.

The discovery of the finite velocity of light revealed that it is
not possible to provide a univocal definition of time with respect
to bodies of reference that are in relative motion. One consequence
of this discovery, as Einstein observes, was to expose the
previously unnoticed interdependence of space and time. That is to
say, prior to the advent of relativity, points in space and instants
of time were taken to be absolute realities -- and time and space
themselves were understood as completely different and independent
things. In the special theory of relativity, Newton's absolutes are
reorganized as follows:
\BP

1. We stipulate that the velocity of light is not only finite, but
that it is an absolute constant -- i.e. that its velocity in a
vacuum is the same for all observers, independent of relative
motion.

2. We stipulate that we will confine our attention to inertial
reference frames -- i.e. to bodies whose rate of relative motion is
uniform.

3. And we continue to presuppose the absolutes of rigid transport
and standard clocks.
\MP

\F
Given these assumptions and qualifications, as Einstein observes,
what has ``physical reality" is neither points in space nor instants
in time, but events, which are understood as specifications of four
numbers in a space-time manifold.  Accordingly, he observes,
\begin{quote}
	There is no absolute (independent of the space of reference)
	relation in space, and no absolute relation in time between two
	events, but there is an absolute (independent of the space of
	reference) relation in space and time, as will appear in the
	sequel. (1922, p.30f.)
\end{quote}
And in the following lecture he is equally clear about the fact that
something like the Newtonian absolute is retained in the special
theory of relativity (Einstein 1922, p.55):
\begin{quote}
	The principle of inertia, in particular, seems to compel us to
	ascribe physically objective properties to the space-time
	continuum. Just as it was consistent from the Newtonian
	standpoint to make both the statements, tempus est absolutum,
	spatium est absolutum, so from the standpoint of the special
	theory of relativity we must say, continuum spatii et temporis
	est absolutum. In this latter statement absolutum means not only
	``physically real," but also ``independent in its physical
	properties, having a physical effect, but not itself influenced
	by physical conditions."
\end{quote}

\F
But it is important that this paragraph occurs on the first page of
the lecture in which Einstein is beginning to introduce the General
Theory of Relativity -- by contrasting it with the Special Theory.
Because the point of this paragraph is that precisely that principle
which seems to compel us to treat the space-time continuum as
physically real and absolute -- precisely that principle is what
loses its privileged status under GR. If we attempt to generalize
the theory of relativity to allow the description of the behavior of
bodies insofar as they are in mutually accelerated reference frames,
we can no longer hope to describe this behavior accurately by means
of coordinate systems referring to rigid reference bodies, nor can we
assume that two clocks in different locations, both of which are at
rest with respect to one reference body will give uniform readings
when considered with respect to another reference body which is in
non-uniform motion with respect to the first.

Our contention, therefore, is that it is only with the advent of GR
that Einstein is forced to fully abandon the Newtonian absolute
(understood in its 4-dimensional Minkowski version). But it is
precisely at this point in the development of Einstein's theory that
one discovers the need for what, in the following section, we call
``the localized absolute."

That is to say, as soon as Einstein attempts to move beyond the
restrictions of the special theory of relativity, two problems occur
at the same time: One is that he is confronted by a need to give a
consistent definition of a local clock. The other is that he needs
to find or create something that can serve as an objective reference
standard (which, as noted above, was one of the main intended --
although not fulfilled -- functions of Newton's notion of absolute
time). But, paradoxical as it may seem, the abandonment of the
rigidity of Minkowskian space-time means that the desired objective
reference standard must be identified, not with some global frame of
reference within which the local system is situated, but with the
local system itself -- insofar as the local system is taken as {\it the}
system.

We resolve these two problems, {\it i.e.} the need for
 a properly defined local clock and an
 objective reference standard,
by introducing the notion of the
local time of a local system on the basis of a total universe,
which is introduced as an objective reference standard.
Our notion of local time, on the one hand, gives
a consistent definition of local systems, each of which can accommodate
a proper clock that serves the requirements of
GR. We associate a local system to each classical point $(t,x)$ of
the 4 dimensional Riemannian manifold $X$,
 so that the local system can accommodate a local clock inside it
which describes the local time at the center of mass of the local system.

On the other hand, we will show
that it is possible to construct
 a concept of a ``total universe" which will serve as the
 ground for an objective reference standard -- without, however,
 violating the strictures of Einstein against an absolute reference
 frame in the classical Newtonian sense.
We represent the total universe as an eigenstate of a Hamiltonian
of infinite degrees of freedom. This allows to define the
local time of each local system with finite degrees of freedom
inside the total universe.
We shall discuss this point in sections
IV and V in detail.

The local system $L$ at a classical point $(t,x)$
 in $X$ has an internal structure
which is {\it independent of}
 the classical mechanical world outside the local system $L$,
as we will see in later sections.
The internal structure of the local system $L$
 is described by quantum mechanics
in our formulation. Since the local system $L$
 is independent of the external
classical world, this assumption of our formulation
 does not lead to any contradiction
as we shall see later. Furthermore, in the local system $L$
a local time  can be defined as a quantum mechanical notion
associated with that local system $L$
 without requiring that time be defined as the problematic
point-specific notion we encountered in Einstein's formulation of
GR. This local time as defined
above -- specific to each local system -- also gives us a local time
for the center of mass of $L$, thus satisfying the requirement of
GR for a definition of local time valid at a particular point.
 Thus we can regard the centers of mass
of various local systems as classical particles obeying GR,
so that we can recover GR in this formulation of
 local systems and local times.

In this way the formulation outlined above
illustrates the possibility of understanding QM and GR as
mutually independent, but at the same time mutually complementary,
such that each supports the other, supplying and supporting features
that cannot be adequately defined within either theory by itself.
QM provides the internal clock which recovers the local clock for GR,
whose realization has been the intrinsic difficulty of GR
from its very beginning as we have seen (see also section V below).

As a basis for this approach, we define,
in the next section, the local clock of a local system
 as a measure of motion inside
the local system. Our notion of local time will give us, as we explain 
below,
a {\it localized} version
of Newton's absolute time.
As we shall see below, by ``localizing" Newtonian time, we will be
enabled to salvage that aspect of Newtonian time which is
presupposed by quantum mechanics, without becoming entangled in
conflict with the theory of relativity.

\vskip20pt

\large
\noindent
{\bf IV. Defining the Local Clock: an alternative notion of time}
\normalsize

\vskip12pt

\noindent
In this section we begin the presentation -- to be continued in part
II of this article -- of an outline of our theory of local time,
which we contend offers a possible way to unify Quantum Mechanics
with General Relativity.

\BP

As noted in Kitada 1994a, p.283, the empirical scientist's notion
of time {\it in actual practice} is necessarily a local one. That is to
say, time does not appear to us until it is measured by some
equipment.
In this respect the observation of time
 is quite different from the observation of
positions and motions, which are perceived directly by our senses.
Even when we use a tool, such as a measuring
stick, to measure the length of a thing, what we
actually do is to look and see which markings on the scale of the
ruler coincide with either extreme of the thing being measured.
The fundamental act of observation here is the perception of this
coincidence (of the ruler-mark with the edge of the thing).
 If we consider how time is measured by means of
clocks, we notice that the measurement process is actually a process
of comparing the motions and positions of certain bodies, so that it
is possible to describe the measurement of time (somewhat
abstractly) as a quotient of certain positions and velocities. With
an analog clock, time is measured by examining the motion of its
hands.
 We look at the hands, and recognize
 that one second passes if the second hand ``moves" one
``scale". We do not measure time directly by our senses,
 but we know time by perceiving the positions and motions of
 the hands of clocks. In this sense time is neither a quantity
 nor a frame given {\it a priori}. What exists first are the positions
 and movements of the bodies relative to our own position.
 The perception of the positions and motions indicates an
 introduction of the common parameter in each system of
 bodies consisting of a finite number of particles. This parameter
 is called time and it is a local notion by nature{\footnote{It
 should be noted here that the foregoing observations concerning
the nature of time measurement are in full agreement with
Einstein, for it is clear that Einstein's understanding of time is
completely ``operational," as may be observed from the fact Einstein
speaks primarily of clocks, not of time as a thing existing in
itself.}}.

       In QM, however, as we noted above in section II,
the notion of time is quite different from that of Einstein.
Obviously, as experimenters, quantum physicists use clocks in the
same way that relativity physicists do, so that in practice they
must implicitly share the understanding of time as a local notion.
But {\it in theory} the Schr\"odinger equation has
traditionally been understood to define the evolution of the
particle states of a system with respect to an externally-given,
``Newtonian," time parameter, which means that for QM the clocks by
which time is measured are understood as if they were completely
external to the system being investigated -- in sharp contrast to
GR, where the fact that the clock is itself an object within the
system is essential to the working of the theory.
 However, in recent years a group
 of mathematical physicists specializing in the study of Schr\"odinger
 wave operators in many-body systems, including Enss and Kitada,
 have made important strides in understanding the asymptotic
 completeness of observables in such systems. Building on
 the work of Enss, Kitada has shown that it is possible to re-interpret
 the role of the time parameter in the Schr\"odinger equation.
 Specifically, it is possible to treat the $t$ in that equation,
 not as an observable, but as a dependent variable, whose specific
 meaning and value are derived, subject to appropriate assumptions,
 from a mathematical consideration based on the velocity and
 momentum of the particles of the system -- which actually are
 observable. The mathematical reasoning which accomplishes this
 result is fairly technical -- it will be summarized below and
 the reader is referred to Kitada 1994a, etc. for a more formal
 presentation of the argument. However, it is possible to state
 briefly here that this analysis, by inverting the customary
 relationship between the time parameter and the observation of
 the velocities and positions of particles as these things
 are interpreted by the Schr\"odinger equation,
 actually brings the understanding of time in QM into a
 much closer and more intelligible relation to that of GR than
 it had before. That is to say, the result of Kitada 1994a is
 to show that time may be understood in QM, no longer as an
 external ``Newtonian" parameter, but as a certain expression
 of the relative motions and positions of the particles making up
 the local system. Thus we show that time may be understood as rooted
 in the internal motions of a local system in QM, just as it is in GR.
 At the same time, however, the theorem of Enss as interpreted and
 extended by Kitada 1994a allows us to explain why this
``local time" must nonetheless appear as if it were an absolute,
 Newtonian time, independent of which point is selected within
 the local system and the same for all. And we will show later
 that it is precisely because it possesses this feature that
 the ``local clock," as we have defined it in the language of QM,
 provides the required logical foundation for the definition of
 local time in GR.

	We are now in a position to define our notion of local time.
 We remark that the following exposition is a rather intuitive
 definition and precise formulation needs some mathematical
 notions and notations as described in sections 4 and 5, pp. 286-288
 of Kitada 1994a. Let $L$ be a local system consisting of
 $N$ number of particles
$1, 2, ..., N$. Then there can
be defined the position vectors $x_1,x_2,\cdots,x_N$
  and momentum vectors $p_1=m_1v_1,p_2=m_2v_2,\cdots,
p_N=m_Nv_N$, where $m_j$  is the mass of the $j$-th particle, so that
the correspondent quantum mechanical selfadjoint operators $X_j=(X_{j1},
X_{j2},X_{j3})$  and $P_j=(P_{j1},P_{j2},P_{j3})$
  in a Hilbert space $\HH=L^2(R^{3n})$ of $N=(n+1)$ particles satisfy
 the so-called canonical commutation relation.
 (This statement is axiom 2
 of Kitada 1994a.) Then the local time $t_L$
 associated with the local system
$L$ is defined as a quotient of position $x_j$  by velocity $v_j=p_j/m_j$
\beq
t_L=\frac{|x_j|}{|v_j|}.
\ene
Here we note that the right hand side of this definition looks
 as if it depends on the number $j$. But
 it is known (Enss 1986)
 that it does not depend on
 $j$, if one defines the right hand side
 as in Kitada 1994a, sections 4-5 (see axiom 3, Theorem 1, and
Definitions 1-3 there, and section V of part II
for more precise descriptions). Thus local time is defined
 as a {\it measure of motion} inside each local system.

        We note that we have defined time only for
 local systems as a parameter of motions, which is abstracted
 from the internal motions inside the local systems. The fact that
 the universe as a whole is not a local system thus makes it
 reasonable to postulate that there is no time associated with
 the total universe. This postulate is axiom 1 of Kitada 1994a.
 This distinction between local systems and the total universe
 is seen more clearly when one notices that the local Hamiltonians
 describing local systems and the total Hamiltonian used to define
 the total universe differ in that the former is
 of finite degrees of freedom while the latter is of infinite degrees
 of freedom, and that the local Hamiltonians are no more than
 convenient approximations to the total Hamiltonian, used instead
 of the true, total Hamiltonian when one observes the outside.
 The fact that each of the local
Hamiltonians of finite degrees of freedom is an approximation, but
only an approximation, of the total Hamiltonian of infinite degrees
of freedom has an interesting and important consequence: Namely it
allows each local system to vary, so that local motions can occur
and local clocks can be defined -- even though the total universe,
consisting of an infinite number of particles is
 stationary{\footnote{The word ``stationary" here is the one used in
 mathematical physics to express an eigenstate of a Hamiltonian as a
 ``stationary state."}}, as we have postulated.

 The stationary nature of
 the total universe will be described in section V of Part II in
 a more precise way. Any local system of a finite number of particles,
 however, can be nonstationary, and can vary inside itself,
 as a consequence of the variation outside the local system,
 which compensates for the change inside the local system,
 so that the stationary nature of the total universe is preserved.

 {\it Because} of the
 fact that the relationship expressed in formula (1) has been
 shown to be independent of the particular choice of particle number,
 this quotient can be understood to hold approximately in the same way
 for any particle in the given local system -- and precisely for
 this reason, this can be understood as defining a common parameter $t_L$
   associated with the local system $L$ itself, rather than with any
 particular point in it, and which we can, therefore, define as the
 ``local time" of the local system. Thus the demonstration, due to Enss,
 that this quotient holds independent of the choice of particle number 
$j$,
 is remarkable in that it not only gives us a way of understanding
 how a definition of time can be derived from the relative motions
 of particles within a local system, but at the same time it shows us
 that this parameter, because it is (approximately) independent
 of the particular choice of particle, can be treated as if it
 actually existed externally, independent of the motions
 on which it is based.

	Nevertheless, once a local time is identified by the formula (1),
 as a measure of motion, as $t=|x|/|v|$ in each local system,
 our definition of local times is a specification or
 a clarification of the `relative, apparent, and common time' measured
`by the means of motion, which is used' `instead of true time'
 in Newton's sense (see the first quotation from {\it Principia},
 Newton 1962). It is a realization of Einstein's
 local nature of time and coordinates, as well, in the sense
 that the local time is defined only for each local system
 consisting of a finite number of particles.

	We note that there is a considerable difference between
 our definition of local times and the conventional understanding of
 the notion of time. The common feature of the conventional
 understanding of time, including Newton's definition of absolute time,
 is that time is something existing or given {\it a priori},
 independently of any of our activities, {\it e.g.} activities
 of observation. In our definition, time is not an {\it a priori}
 existence, but a convenient measure of motions inside each local system.
 Our definition of local times mentioned above is that a local time is
 a clock -- which measures, not time, but the motions of the local system.
 Unlike the conventional understanding where time is given {\it a priori},
 the local clock does not measure time, but it {\it is} time.
 Further, as we will state in section V, the proper clock is the local
 system itself, and it is a necessary manifestation of that local system.
 In this sense, ``clocking" is the natural activity of any local system.
 It follows from this that to be an existing thing in the world
 necessarily involves clocking, without which there is no interaction.
 In these respects, our position is in complete opposition to the
 conventional understanding of time measurement, where time is given
{\it a priori} and clocks measure those times, therefore the measurement
 of time is an incidental activity. Contrary to the conventional
 understanding, our view is that all beings are engaged in measuring
 and observing, and the activities of measuring and observing are not
 incidental, but pertain to the essence of all interactions. If we
 are permitted to express it somewhat boldly, we have turned things
 completely around: It is not that things exist and their duration
 is incidentally expressed by clocks. According to our formulation,
 clocks exist and their operation is necessarily expressed by duration.

Philosophically speaking, our understanding stated in axiom 1
about the totality of
nature reflects that of Spinoza, especially insofar as Spinoza says
that the totality of nature is Eternal, and defines Eternity as
follows (Spinoza, Ethics, Part I, Definition 8, from  E. Curley 1985
p.409):

\begin{quote}

	D8: By eternity I understand existence itself, insofar as it is
	conceived to follow necessarily from the definition alone of the
	eternal thing.

	Explanation: For such existence, like the essence of a thing,
	is conceived as an eternal truth, and on that account cannot be
	explained by duration or time, even if the duration is conceived
	to be without beginning or end.

\end{quote}

\noindent
Our axiom 1 which asserts that the total universe, which will be denoted
 $\phi$, is stationary means in its mathematical formulation that
 it is an eigenstate of a total Hamiltonian  $H$.
 This means that the universe $\phi$
  is an eternal truth, which cannot be explained in terms of
 duration or time. In fact, the eigenstate in itself contains
 no reference to time, as may be seen from its definition:
  $H\phi=\lambda\phi$ for some real number $\lambda$.
The reader might think that this definition just states
that the entire universe $\phi$ is frozen at an instant which lasts
 forever without a beginning or end.
 However, as we will see, the total universe $\phi$
 has infinite degrees of freedom inside itself, as internal motion of
 finite and local systems, and never freezes. Therefore,
as an existence itself, the universe $\phi$ does not change, however,
at the same time, it is not frozen internally.
These two seemingly contradictory aspects of the universe
$\phi$ are possible by virtue of the quantum mechanical nature of
the definition of eigenstates.

	To sum up, the universe itself does not change. However,
inside itself, the universe can vary quantum mechanically,
 in any local region or in any local system consisting of a finite
number of (quantum mechanical) particles. Therefore, we can define a
 {\it local time} in each local system as a measure or a clock of
 (quantum mechanical) motions in that local system.

Let us consider, finally, the relationship that our notion of time
bears to those of Newton and Einstein. First, we are in agreement
with Einstein in abandoning the Newtonian conception of time as an
absolute time pertaining to the entire universe. We have defined
local times for describing local motions in a way that will be shown
to be consistent with GR.  However, we have also preserved certain
aspects of the Newtonian conception: we have localized Newtonian
time by showing that the local clock, as we have defined it, gives
an approximation to Newtonian time which is valid for any particle
of the local system. And, in another sense, we have retained for the
universe as a whole the absoluteness of Newton -- but without the
flow of time -- since our definition of local time involves the
consequence (whose implications will be explored in Part II) that
the universe itself is not in time, but is eternal, as Spinoza has
defined that term.

\vskip24pt

\vskip38pt

\large
\noindent
{\bf References}

\BP

\small

\F
E. Curley, 1985, The Collected Works of Spinoza,
Volume I, Edited and Translated by Edwin Curley,
 Princeton University Press, Princeton, New Jersey.

\F
A. Einstein, 1905, ``Zur Elektrodynamik bewegter K\"orper,"
Annalen der Physik, {\bf 17}, 891-921.

\F
A. Einstein, 1916, ``Die Grundlage der algemeinen Relativit\"atstheorie,"
 Annalen der Physik, Ser.4, {\bf 49}, 769-822.

\F
A. Einstein, 1920, Relativity, The Special \& The General Theory,
 Tr. R. W. Lawson, Methuen \& Co. LTD.

\F
A. Einstein, 1922, The Meaning of Relativity, (Stafford Little Lectures of
 Princeton University, presented in May, 1921), 5th ed.,
Princeton University Press, 1956.

\F
V. Enss, 1986,  ``Introduction  to  asymptotic observables for
 multiparticle quantum scattering," in Schr\"odinger Operators,
Aarhus 1985, ed. E. Balslev, Lect. Note in Math. {\bf 1218}, 
Springer-Verlag,
pp.61-92.

\F
R. P. Feynman, 1948, ``Space-time approach to non-relativistic quantum
 mechanics," Rev. Modern Phys., {\bf 20}, 367-387.

\F
J. B. Hartle, 1993, ``The spacetime approach to quantum mechanics," 
in Proceedings of the International Symposium on Quantum Physics and 
the Universe, Waseda University, Tokyo, Japan, August 23-27, 
1992. (gr-qc/9210004)

\F
W. Heisenberg, 1925, ``\"Uber quantentheoretische Umdeutung kinematischer
 und mechanischer Beziehungen," Zeitschrift f\"ur Physik, {\bf 33}, 
879-893.

\F
C. J. Isham, 1993,  ``Canonical quantum gravity and the problem of time,"
 in Proceedings of the NATO Advanced Study Institute, Salamanca, 
June 1992, Kluwer Academic Publishers.  (gr-qc/9210011)

\F
H. Kitada, 1980, ``On a construction of the fundamental solution
for Schr\"odinger equations," J. Fac. Sci. Univ. Tokyo, Sec. IA,
{\bf 27}, 193-226.

\F
H. Kitada, 1994a,  ``Theory of local times," Il Nuovo Cimento,
{\bf 109 B, N. 3}, 281-302. (astro-ph/9309051)

\F
H. Kitada, 1994b, ``Theory of local times II. Another formulation and
examples," preprint. (gr-qc/9403007)

\F
H. Kitada, L. R. Fletcher, Local Time and the Unification of Physics, Part 
II, to appear in Apeiron.

\F
I. Newton, 1962, Sir Isaac Newton Principia, Vol. I The Motion of Bodies,
 Motte's translation Revised by Cajori, Tr. Andrew Motte ed.
Florian Cajori, Univ. of California Press, Berkeley, Los Angeles, London.

\F
E. Schr\"odinger, 1926, ``Quantisierung als Eigenwertproblem 1, 2,"
 Annalen der Physik, {\bf 79}, 361-376, 489-527.

\F
W. G. Unruh, 1993, ``Time, gravity, and quantum mechanics," preprint. 
(gr-qc/9312027)

\F
H. Weyl, 1952,  Space -- Time -- Matter, Translated by H. L. Brose,
 Dover Publ., Inc. New York.

\end{document}